# Error Correction Capacity of Unary Coding

Pushpa Sree Potluri[1]

**Abstract**
Unary coding has found applications in data compression, neural network training, and in explaining the production mechanism of birdsong. Unary coding is redundant; therefore it should have inherent error correction capacity. An expression for the error correction capability of unary coding for the correction of single errors has been derived in this paper.

1. **Introduction**

The unary number system is the base-1 system. It is the simplest number system to represent natural numbers. The unary code of a number *n* is represented by *n* ones followed by a zero or by *n* zero bits followed by 1 bit [1]. Unary codes have found applications in data compression [2],[3], neural network training [4]-[11], and biology in the study of avian birdsong production [12]-14]. One can also claim that the additivity of physics is somewhat like the tallying of unary coding [15],[16]. Unary coding has also been seen as the precursor to the development of number systems [17].

Some representations of unary number system use *n-1* ones followed by a zero or with the corresponding number of zeroes followed by a one. Here we use the mapping of the left column of Table 1.

Table 1. An example of the unary code

| N | Unary code | Alternative code |
|---|---|---|
| 0 | 0 | 0 |
| 1 | 10 | 01 |
| 2 | 110 | 001 |
| 3 | 1110 | 0001 |
| 4 | 11110 | 00001 |
| 5 | 111110 | 000001 |
| 6 | 1111110 | 0000001 |
| 7 | 11111110 | 00000001 |
| 8 | 111111110 | 000000001 |
| 9 | 1111111110 | 0000000001 |
| 10 | 11111111110 | 00000000001 |

The unary number system may also be seen as a space coding of numerical information where the location determines the value of the number. In order to denote a specific value, corresponding slot should be marked. This is shown in Figure 1.

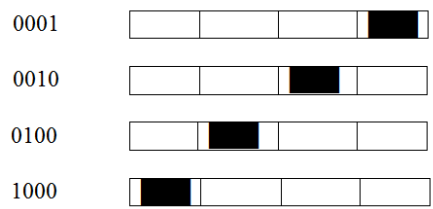

Figure 1. Space representation of numbers

[1] Oklahoma State University, Stillwater, OK 74078

If this coding is further modified by the rule that all places to the left of 1 should be replaced by 1, we get the familiar:

    1: 1111
    2: 1110
    3: 1100
    4: 1000

Unary coding is most useful in counting or tallying ongoing results. Unary coding is used as a part of some data compression algorithms such as Golomb coding. In the Golomb code, unary coding is used to encode the quotient part of the Golomb code word.

This paper reviews some applications of unary coding and then presents an error-correction analysis of the codes.

## 2. Previous Research

*Optimal coding.* In the Golomb code [2], numbers are divided into groups of equal size *m*. It is denoted as Golomb-m. The code word is represented as follows

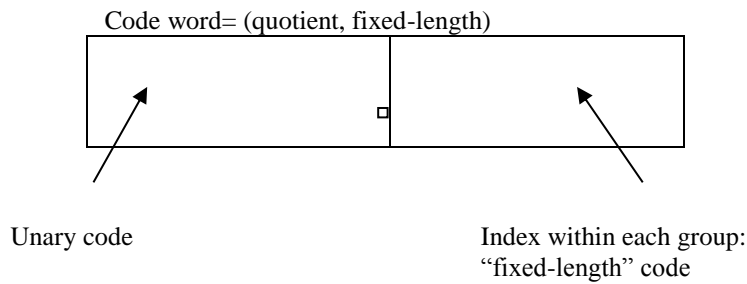

Code word= (quotient, fixed-length)

Unary code                                       Index within each group: "fixed-length" code

For example, let us consider m=8. The Golomb-8 codes are given as in Table 2.

Table 2. Example of Golomb code

| N | Q | R | Code word |
|---|---|---|---|
| 8 | 1 | 0 | **10**000 |
| 9 | 1 | 1 | **10**001 |
| 10 | 1 | 2 | **10**010 |
| 11 | 1 | 3 | **10**011 |
| 12 | 1 | 4 | **10**100 |
| 13 | 1 | 5 | **10**101 |

A unary code is optimal if it satisfies the following conditions:
    The data in which the numbers are positive integers and the probability of lower number is greater than or equal to probability of higher number

$$p(1) \geq p(2) \geq p(3) \ldots \geq \ldots$$

Then the code length ($\lambda$) should be $\lambda_1 \leq \lambda_2 \leq \lambda_3 \leq \ldots$.
The unary code for *i* is *i−1* 0's, followed by a 1, where $\lambda_i = i$.

The probability *p(i)* such that $\lambda_i = log(1/p(i))$ that is, if $p(i)=2^{-i}$ then the average code length would be equal to the entropy.

*Neural network learning.* Fixed length unary coding was used in instantaneously trained neural networks [4]-[11] to ensure that learning a specific point makes it possible to learn all adjacent (in the Hamming distance sense) points .

The corner classification network is used to find the weights to the hidden layer. In CC4, a feed forward network architecture consisting of three layers of neurons (input layer, hidden layer and output layer) is used. The number of input neurons is equal to the length of input patterns or vector plus one. The number of hidden neurons is equal to number of training samples and each hidden neuron corresponds to one training example.

If input neuron receives 1, its weight to the hidden neuron is 1 or else it is -1. To provide for effective non-zero thresholds to the hyper plane realized by the node, we assume an extra input $x_{n+1}$. The weight between the biased neuron and the hidden neuron is (*r-s*+1), where *r* is the radius of generalization and '*s*' is the number of ones in the input sequence. The weights in the output layer are equal to 1 if the output value is 1 and -1 if the output value is 0.

$$W_i[j] = \begin{cases} 1 & \text{if } x_i[j] = 1 \\ -1 & \text{if } x_i[j] = 0 \\ r\text{-}s+1 & \text{if } j = n. \end{cases}$$

The choice of *r* will depend on the nature of generalization. If no generalization is needed, r=0. For exemplar patterns choice of *r* defines the degree of error correction and it will also depend on the number of training samples. Since the weights are 1,-1 or 0 it is clear that actual competitions are minimal. In the general case, the only weight that can be greater in magnitude than 1 is the one associated with the biased neuron.

*Applications in Biology.* Birdsong is an important model system in behavioral neurobiology, as it offers both easily measured sensory system and motor patterns and a discrete nucleus effector system. The HVC (High level Vocal center nucleus) has several classes of neurons with apparently specialized roles in song production necessary for both the learning and the production of the birdsong [18]-[20].

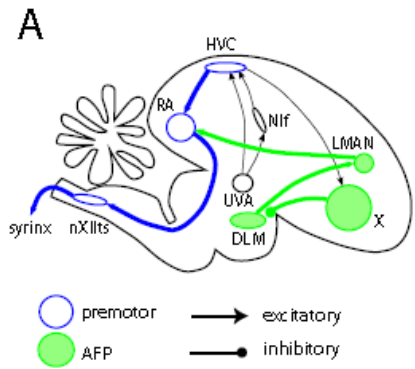

Figure 2. Song learning pathway in the birds *(I.R.Fiete, H.S.Seung [18])*

Birdsong learning involves a group of distinct brain areas that are aligned in two connecting pathways. Vocal learning involves anterior forebrain pathway (AFP), composed of area X, which is a homologue to basal ganglia to the dorso-lateral division of the medial thalamus (DLM) to the lateral part of the magnocellular nucleus of anterior nidopallium (LMAN). Posterior descending pathway (Vocal production), composed of HVC, projects to the robust nucleus of the arcopallium (RA) to the tracheosyringical part of the hypoglassal nucleus(nXllts). The posterior descending pathway is required throughout a bird's life for normal song production, while the anterior forebrain pathway(AFP) is necessary for song learning. Both neural pathways in the song system begin at the level of HVC.

In the synaptic chain model, the neurons in a network are divided into groups, and the groups are ordered in a sequence. From each group of neurons, there are excitatory synaptic connections to the neurons in the next group of the sequence. The groups and the synaptic connections are like the links of a chain.

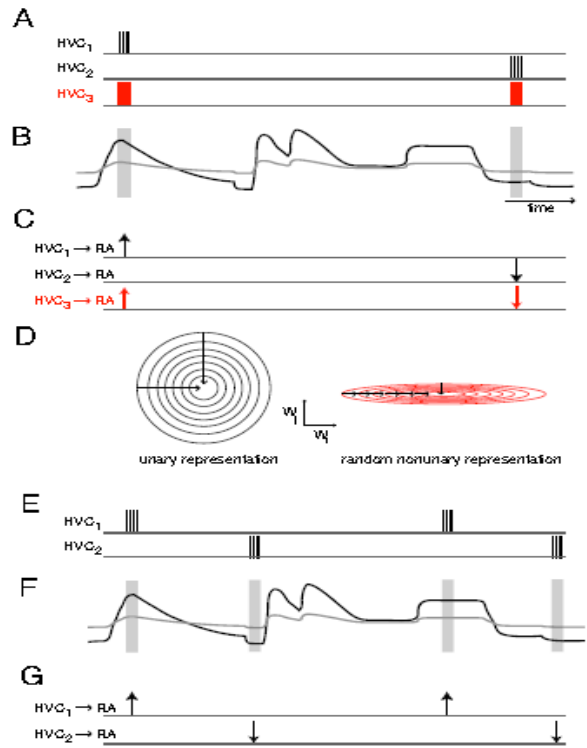

Figure 3. Unary coding in HVC *(I.R.Fiete, H.S.Seung [18])*

The fact that the impulses are generated from different parts of HVC is in itself evidence of the unary coding of the information.

### 3. Derivation of error correction formula

Since there is much redundancy in the unary code, it is to be expected that it could be able to correct some errors. Here we consider single errors, of which a subset would be easiest to correct.

We consider the binary symmetric channel. The bit error probability is p. The probability of 0/1 returning as 0/1 after transmission is *(1-p)* i.e. no error condition and the probability of 0/1 returning as 1/0 is *p*.

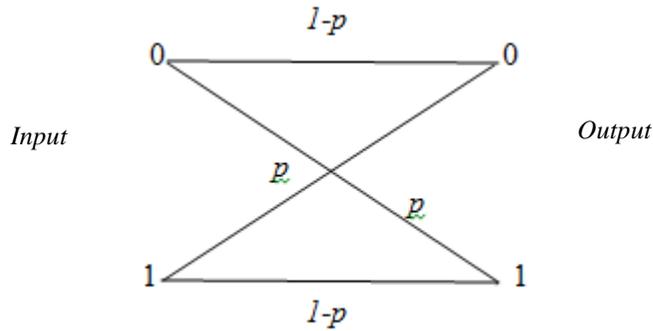

Figure 4. Binary symmetric channel (bit error probability is p)

**Theorem:** The number of single errors that can be corrected in the unary code of number *n* is $(n-1)^2$.

**Proof**: The total number of single errors that can be corrected would vary across numbers. For example, for the number 0, all errors excepting at the right-most place will be corrected, which is a total of n-1; same for the all 1 sequence. For the number 1, single errors excepting at the right-most two places will be corrected; for number 2, it likewise gives us a similar number of correctible errors. In other words, for the general case, n-3 cases are correctible for each number. Gathering all these terms, we get the number of sequences out of the total number that can be corrected:

$$= 2(n-1) + (n-1)(n-3)$$
$$= (n-1)^2$$

Example. Consider the unary code for *n=5*

Table 3. Correctible single errors for n=5

| Code Word | Total number of single errors | Errors Corrected | Generalized Form |
|---|---|---|---|
| 00000 | 00001, 00010, 00100, 01000, 10000 | 10000, 01000, 00100, 00010 | (n-1) |
| 00001 | 00000, 00011, 00101, 01001, 10001 | 10001, 01001 | (n-3) |
| 00011 | 00010, 00001, 00111, 01011, 10011 | 10011, 01011 | (n-3) |
| 00111 | 00110, 00101, 00011, 01111, 10111 | 00101, 00110 | (n-3) |
| 01111 | 01110, 01101, 01011, 00111, 11111 | 01101, 01110 | (n-3) |
| 11111 | 11110, 11101, 11011, 10111, 01111 | 10111, 11011, 11101, 11110 | (n-1) |

Thus the number of single errors that can be corrected for *n=5* are $(5-1)^2=16$

As we are considering the single errors, only one bit of the string returns as the wrong bit after transmission. Hence, the probability of error correction is represented as

$$P(correction) = (n-1)^2 p(1-p)^{n-1}$$

Given that the total of single errors that can take place is (n+1)n, the correction capacity for single errors is;

$$P(CorrectionCapacity) = \frac{(n-1)^2 p(1-p)^{n-1}}{n(n+1)}$$

The value of *p* at which the maximum error correction occurs is by finding the derivative of the expression and putting that equal to zero. We get:

$$-p(n-1)(1-p)^{n-2} + (1-p)^{n-1} = 0$$

This yields *p=1/n*.

Figure 5 plots some examples of the error correction probability.

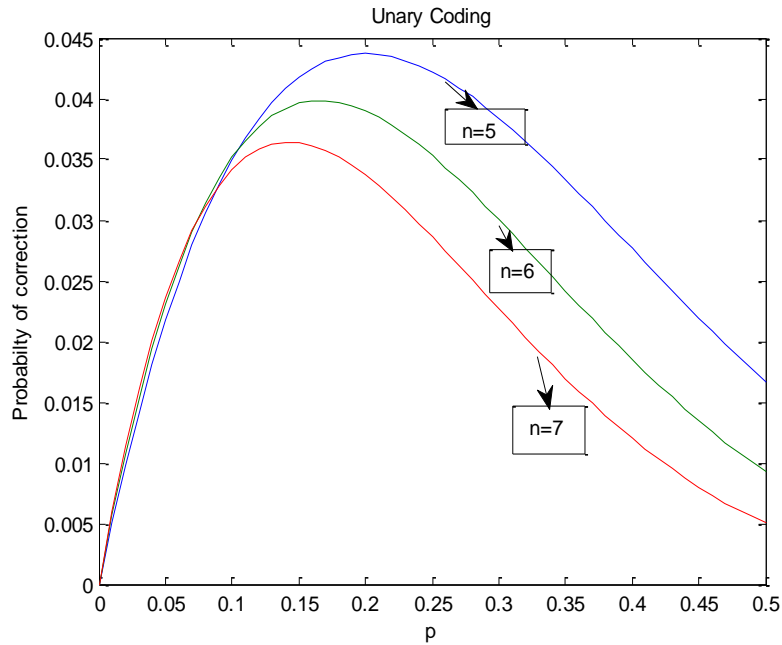

Figure 5. Three examples of single error correction probability

**Double errors**
Here we consider the example of double errors for the case n=5. The number of double errors that can be corrected for *n=5*

Table 4. Double error correction for n=5

| Code word | Errors corrected |
| --- | --- |
| 00000 | 11000, 01100, 10100, 01010, 10010 |
| 00001 | 11001, 10101 |
| 00011 |  |
| 00111 |  |
| 01111 |  |
| 11111 | 11100, 11010, 10110 |

Total number of double errors that can be corrected for *n=5* are 10. On the other hand, the total number of double errors that will take place for each row of Table 4 is $^5C_2 = 10$. Thus the capacity of double error correction in this case is:

$$P(double\ error\ correction) = \frac{1}{6} 10\ p^2(1-p)^3$$

**Conclusions**

We have found an expression for intrinsic error correction capacity of unary coding. Since unary coding has a variety of applications ranging from data compression to neural network training, it may be possible to exploit this capacity in these application. Furthermore, the intrinsic error correction may be a reason why unary coding is used in biological networks.